\documentstyle[12pt,epsfig]{article}
\newcommand\beq{\begin{equation}}
\newcommand\eeq{\end{equation}}
\newcommand\bea{\begin{eqnarray}}
\newcommand\eea{\end{eqnarray}}

\begin{document}

\begin{center}
{\Large Magnetization of $Mn_{12} Ac$ in a slowly varying magnetic field: 
an {\it ab initio} study}
\end{center}

\vskip .5 true cm
\centerline{\bf Indranil Rudra$^1$, S. Ramasesha$^1$ and Diptiman Sen$^2$} 
\vskip .5 true cm

\centerline{\it $^1$ Solid State and Structural Chemistry Unit} 
\centerline{\it $^2$ Centre for Theoretical Studies}  
\centerline{\it Indian Institute of Science, Bangalore 560012, India} 
\vskip 1.5 true cm

\begin{abstract}

Beginning with a Heisenberg spin Hamiltonian for the manganese ions in the 
$Mn_{12} Ac$ molecule, we find a number of low-energy states of the system. We 
use these states to solve the time-dependent Schr$\ddot o$dinger equation and 
find the magnetization of the molecule in the presence of a slowly varying 
magnetic field. We study the effects of the field sweep rate, fourth order 
anisotropic spin interactions and a transverse field on the weights of the 
different states as well as the magnetization steps which are known to occur 
in the hysteresis plots in this system. We find that the fourth order term 
and a slow field sweep rate are crucial for obtaining prominent steps in 
magnetization in the hysteresis plots.

\end{abstract}
\vskip .5 true cm

~~~~~~ PACS number: ~75.45.+j, ~75.60.Ej

\newpage

\section{\bf Introduction}

Recently some magnetic systems have been discovered in which simple quantum 
mechanical principles lead to some striking macroscopic phenomena. In 
particular, the observation of discrete steps in the magnetization of 
single crystals of the compound $Mn_{12}Ac$ in the presence of a 
time-dependent magnetic field has evoked considerable theoretical interest. 
This has been termed variously as quantum hysteresis and quantum or resonance 
tunneling of magnetization. 

The basic underlying physics is easy to understand 
\cite{chudnovsky}. The system consists of magnetic molecules which interact 
only weakly with each other. Each molecule is a cluster consisting of a core 
tetrahedron of four $Mn^{4+}$ ions each with a spin of ${3 \over 2}$, and an 
outer crown consisting of eight $Mn^{3+}$ ions each with spin $2$. The 
intracore as well as intracrown magnetic interactions are ferromagnetic, 
while the interactions between the spins in the core and the crown is 
antiferromagnetic (see Fig. 1). Thus each molecule is a ferrimagnetic cluster
with a ground state spin of $10$. In the cluster, the dipolar interaction
between the spins leaves only the $M_s=10$ and $-10$ states degenerate. The
application of a magnetic field lifts this degeneracy, resulting in a nonzero
magnetization. As the field is increased, different pairs of $M_s$ states 
become degenerate at certain values of the field. At those particular 
fields, the presence of matrix elements between the degenerate states, 
provided either by a weak transverse component in the 
applied magnetic field or by higher order spin-spin interactions, causes 
tunneling between the states. This results in a jump in the magnetization. 
At all other values of the field where there are no energy degeneracies, the 
plot of magnetization {\it vs} field shows plateaus or discrete steps provided 
the sweep rate of the magnetic field is not too low \cite{perenboom}. This is 
because, according to the Landau-Zener theory \cite{zener}, the tunneling 
amplitude to go from one magnetization state to another is very small unless 
the sweeping frequency is so low that it is comparable to the matrix element 
connecting the two states. Such steps in the magnetization are also seen 
in another magnetic cluster $Fe_8$ \cite{sangregorio}, although the effect is 
less dramatic there than in the $Mn_{12}$ cluster.

In recent years, there have been many model calculations which illustrate such 
steps in the $M$ {\it vs} $H$ curves \cite{katsnelson,raedt1}. These models 
involve either the presence of a transverse magnetic field, or higher order 
spin couplings which leads to a term of the type $c (\hat{S}_x^4+\hat{S}_y^4)$
allowed by the symmetry of the cluster. However, most of these calculations 
have been restricted to the spin-$10$ manifold, {\it i.e.}, a total of $21$
states. In contrast, we have carried out an explicit calculation of the 
low-lying states of a $Mn_{12} Ac$ cluster using a Heisenberg spin model; 
we then consider $117$ of these states which involve several 
different values of the total spin. We have studied quantum tunneling between
these low-lying states by setting up a Hamiltonian in this subspace 
of states which includes, besides the multipolar spin-spin interactions and 
a transverse magnetic field, different gyromagnetic ratios for the core and
crown spins. This last interaction, which is reasonable to introduce because
of the different environments around the core and crown spins, has the effect
of mixing up the different spin states of the cluster when a magnetic field
is applied; it is therefore essential to keep states with different values 
of the total spin in the calculation, rather than restrict oneself to the 
spin-$10$ ground state manifold. We have then evolved an initial state, which 
is taken to be the ground state with a specific value of $M_s = S_z$ 
in the absence of the magnetic field, by using the time-dependent formulation 
of the problem in the restricted subspace. In the following section, we 
formulate the problem and present some numerical details. This will be 
followed by a section on the results of our time evolution studies and a 
discussion of the results.

\section{\bf Theoretical Model and Computational Details}

The dimension of the Fock space spanned by the magnetic cluster 
$Mn_{12}Ac$ is exactly a hundred million. When specialized to the states of 
interest in the low energy sector, the dimensionalities reduce considerably 
even though they are still large. Table 1 gives the dimensionalities of the 
Hilbert spaces corresponding to the various $M_s$ values of interest. These 
states can be represented by $32$ bit integers by associating 
two bits with each of the core spins of $s={3 \over 2}$ (namely, $M_s = - {3 
\over 2} \rightarrow 00, M_{s} = - {1 \over 2} \rightarrow 01, M_s = +{1 \over 
2} \rightarrow 10$, and $M_s = +{3 \over 2} \rightarrow 11$),
and three bits with each of the outer spins with $s=2$ ($M_s = - 2 \rightarrow
000, M_s = - 1 \rightarrow 001, M_s = 0 \rightarrow 010, M_s = +1 \rightarrow
011$ and $M_s = +2 \rightarrow 100$); these states are generated in an 
ascending order of the $32$ bit integers that represent them.

The Hamiltonian describing the spin interactions in the cluster is given by
\beq
\hat{H} = -J \sum_{<ij>_{core}} \hat{S}_i\cdot\hat{S}_j - J \sum_{<ij>_{crown}} 
\hat{S}_i\cdot\hat{S}_j + J^{\prime} \sum_{<i_{core}j_{crown}>}\hat{S}_i
\cdot\hat{S}_j ~,
\label{ham1}
\eeq
where $J$ is ferromagnetic, $J^{\prime}$ is antiferromagnetic, and the 
summations are taken over nearest neighbors. We have taken $J=1.0$ and 
$J^{\prime} =0.2$ in our calculations. Every crown spin has two nearest 
neighbors in the crown while every core spin has three nearest neighbors in 
the core corresponding to the tetrahedral geometry of the core spin. Every 
core spin also has three nearest neighbor crown spins.

The Hamiltonian matrix is set-up in the desired $M_s$ space, which in our case
is restricted to $M_s =9, 10$ and $11$. In each subspace we have obtained a 
few low-lying states using the Davidson algorithm \cite{davidson}. 
We have also calculated the 
spin densities and the spin-spin correlation functions in each of
the states. Using the spin-spin correlation functions, we have computed the
expectation value of $S^2_{total}$ operator, from which we have identified the
total spin of the state. We also compute the total spin density of the core 
and crown spins in each of these states. From the total spin and the spin 
density of the core and crown spins in the given $M_s$ state, we have used 
the spin ladder operators to obtain the spin densities in the
core and crown of the cluster in all the allowed
$M_s$ states. These are later used in computing the magnetization response of 
the system. We find that the lowest five states with the above values of $M_s$ 
belong to $S_{total} = 10, 11, 12, 10$ and $13$; their energies are given in 
Table 2. We therefore have a total of $117$ low-lying states. 

To study quantum tunneling we have considered all these low-lying 
states and the following Hamiltonian, 
\beq
\hat{H} = E_s - D~\hat{S}_{z,total}^2 + c~(\hat{S}_{x,total}^4 + 
\hat{S}_{y,total}^4) - g_{crown}~h(t)~\hat{S}_{z,crown} - g_{core }~h(t)~
\hat{S}_{z,core} ~.
\label{ham2}
\eeq
Here $D$ is the quadratic anisotropy factor, $g_{crown}$ and $g_{core}$ are 
the Land\'e $g$-factors for the crown and core spin respectively, and $h(t)$ 
is the time-dependent magnetic field. We have chosen $D = 10^{-3}$
and $c = 10^{-3}$ (in units of J) in accordance with 
the experimental values \cite{friedman1,hernandez}. We take $g_{crown} =1.85$
and $g_{core} =2.0$ in order to keep the average $g$ value equal to the 
experimental value of $1.9$. The constants $E_s$ in (\ref{ham2}) correspond to 
the five lowest energies obtained from Eq. (\ref{ham1}) and given in Table 2. 
The fourth order anisotropy term allows transition between states with $\Delta 
M_s = \pm 4$. To study the evolution of the magnetization as a function of the 
applied magnetic field, we start from an axial field equal to $-1$ (in
units of $J/ \hbar$) and then 
slowly increase it in steps till it equals $+1$; the exact
procedure for sweeping the field is described below.

We have studied the time evolution of the system by solving the time-dependent 
Schr$\ddot o$dinger equation
\beq
i\hbar \frac{d\psi}{dt} = \hat{H}(t) \psi ~.
\eeq
We assume the system to be in the state with $S=10,~M_s=-10$ (all-spins-down 
state) at time $t=0$. This is the initial state which is time evolved 
according to the equation
\beq
\psi(t+\delta t) = e^{-i \hat{H}(t+\frac{\delta t}{2}) \delta t / \hbar} ~ 
\psi(t)~.
\eeq
The evolution is carried out by explicit diagonalization of the Hamiltonian
matrix ${\bf H}(t+\frac{\delta t}{2})$, and using the resulting eigenvalues 
and eigenvectors to evaluate the matrix of the time evolution operator $e^{-i 
\hat{H}(t+\frac{\delta t}{2}) \delta t/\hbar}$. Since the Hamiltonian matrix 
is in a truncated basis of $117$ eigenstates of the magnetic cluster, it is 
possible to repeatedly carry out the time evolution in small time steps of 
size $\Delta t$. 

\section{\bf Results and Discussion}

We have carried out a systematic investigation of the dependence of the 
magnetization steps on the field sweep rate, the fourth order anisotropy 
term, and the presence of a transverse field. Before doing that, it is
useful to have an idea of the energy levels of the system as a function
of the magnetic field.

Fig. 2 shows the the energy levels of the Hamiltonian in Eq. (\ref{ham2}) 
with a constant axial magnetic field $h(t) =H_z$. We can 
see from that figure that there are at least three values of the
axial field where level crossings occur. In the case of 
interacting spins with exchange constant $J$ and field value $H_z = 0$, levels 
with opposite magnetizations $\pm M_s$ are degenerate. But when we 
introduce quantum fluctuations, such as the fourth order anisotropy term, then 
the magnetization is no longer conserved. Crossings occur around $H_z = \pm 
0.25$, $\pm 0.4$ and $\pm 0.5$ (in units of $J/\hbar$). The energy spectrum 
is symmetric about $H_z 
= 0$. We should expect to see jumps in the magnetization value at those values 
of the field where the crossings occur. Besides, as we will see, the 
occurrence and widths of plateaus in the magnetization are strongly dependent 
on the field sweep rate. The number of plateaus and their locations and widths 
depend on the probability of tunneling from one magnetization state to 
another. This probability increases when the time scale of sweeping
matches with the time scale of tunneling. In that case, the probability 
of staying in the same eigenstate is small; the state is scattered 
into another eigenstate which produces a step in the magnetization plot. 

Following the technique used by De Raedt {\it et al} \cite{raedt1}, we now 
study the behavior of the magnetization as the field is changed with time. The 
magnetic field $H_z$ is increased from $-1$ to $1$ in steps of $0.02$. At each 
value of the field, the state is evolved for $300$ time steps of size $\Delta 
t$. We have considered two different time step values given by $\Delta t = 
0.01 ~\hbar /J$ and $0.1 ~\hbar /J$; thus each value of the field is kept 
fixed for a time equal to $3 ~\hbar/J$ and $30 ~\hbar/J$ respectively. The 
field sweep rate is given by $0.02/(300 \Delta t)$; we therefore have two 
sweep rates differing by a factor of $10$. At each time step, the time evolved 
state is used to compute the magnetization $M$ given by
\beq
M(t)= \langle \psi(t) | \hat{S}_{z,total} |\psi(t) \rangle ~.
\eeq
The magnetization at each value of the field is then taken to be the average 
of the magnetization computed over all the time steps for which the field is
held fixed.

In Fig. 3, we show the step behavior of the magnetization with the 
applied field. The upper curve is for a time step equal to $0.1$ (in units of 
$\hbar /J$), while the lower curve is for a time step of $0.01$.
We observe jumps and steps in the magnetization plot at field values 
of approximately $H_z =0.25, 0.4$ and $0.5$. Before the first jump in the 
magnetization the magnetization value remains almost constant at $-10$. The 
reason why we do not see any jumps at $H_z = -0.25, -0.4$ or $-0.5$ is because 
of the significant energy difference between the spin-$10$ ground state and 
the higher spin
excited states; this makes the scattering probability extremely small.
We observe a remarkable thing that the magnetization value seems to saturate 
after a certain time evolution, but it never approaches the state with all 
spins up. We can argue that in our model the system can only gain or loose
energy by interacting with the time-dependent field, and there is no 
interaction with the environment through, for example, spin-phonon or nuclear 
spin-electron spin interactions. So even for very large $H_z$ the 
magnetization does not reach the saturation value in a finite time. 
However, the magnetization
does reach a higher value for large fields if the sweep rate is lower, since 
there is more time to tunnel to the lowest energy states in that case. The 
inset of Fig. 3 shows the result obtained when the field is held fixed for a 
longer time equal to $300 ~\hbar /J$ corresponding to $3000$ time steps 
of size $0.1 ~\hbar /J$ each. Note that the plateau in the inset occurs at a 
different value of the magnetization compared to the plateaus in the two 
curves in the main figure where the sweep rates were faster. This is due to
tunneling to nearly degenerate states with different values of the 
magnetization.

Experimentally we see transitions between states for which $\Delta m 
= \pm 1$ \cite{hernandez}. This is because, even in a single crystal
of $Mn_{12}Ac$, all the clusters do not have exactly the same orientation.
This would imply that the magnetic field seen by a cluster would also have
a transverse component due to a slight misalignment of the axial field. 
To account for this fact we have added a term $-g H_x \hat{S}_x$ to the 
Hamiltonian in Eq. (\ref{ham2}). In Fig. 4, we show the magnetization plateaus 
in the presence of a fixed and small transverse field equal to $0.01 ~J/\hbar$
when the axial field is applied in the same manner as before. 
For the lower sweep rate, we do see one more plateau here than in 
Fig. 3 where there was only a fourth order term and no transverse field.

In Fig. 5, we show the magnetization when there is only a transverse field
equal to $0.05$ and no fourth order term. On comparing this with Fig. 4, we 
can see that in the presence of a higher transverse field the magnetization is 
much larger for positive values of the axial field; in fact, the magnetization
almost reaches the maximum possible value of $10$ for the lower sweep
rate. This is because excited states start getting populated more easily 
when the magnitude of the leaking term is larger. We also note that there 
are no intermediate plateaus for the lower sweep rate in Fig. 5 because the 
system has more time to evolve to the lowest energy state at each field.

It is instructive to contrast the effects of the fourth order term vs. a 
transverse field on the dispersion of the low-lying energy levels. In Fig. 6, 
we show the energy levels for two different axial fields equal to $0$ and 
$0.26$. The two figures on the left have the fourth term but no transverse 
field, while the two figures on the right have a transverse field of $0.05$ 
but no fourth order term. 
(For ease of comparison, we have subtracted out appropriate constants 
to make the ground state energy equal to zero in all the four figures). It is 
clear that the energy levels have a much greater dispersion in the presence of 
the fourth order term; this means that the gaps are larger, the tunneling 
amplitudes will be smaller and the magnetization plateaus would be more 
prominent in that case. With a transverse field present, the energy levels
almost form a continuum with very small gaps; hence the system can easily
make transitions from one state to another and the plateaus would be smaller.
This emphasizes that the fourth order term is essential for observing 
magnetization plateaus, and explains why there are more plateaus in Fig. 3 
than in Fig. 5.

To study the populations of different $M_s$ states at the different plateau
regions, we show histograms in Figs. 7 (a) and (b). $|C|^2$ is the sum
of the squares of the modulus of the coefficients for each value 
of $M_s$ in the time evolved 
state $\psi(t)$ averaged over the time for which the field is held constant. 
We have followed the changes in $|C|^2$ for different $M_s$ states at 
three values of the axial fields where the plateaus begin in Fig. 3. In Fig. 
7 (a), we show the change of weights of different $M_s$ values at the 
three onset values of the field. We can clearly see 
that the $\hat{S}_x^4 + \hat{S}_y^4$ term connects states with $\Delta M_s = 
\pm 4$. This plot also explains why the saturation magnetization value never 
reaches $+10$. The $M_s = -10$ state has a considerable weight in all three 
plateaus although it decreases gradually at higher magnetic fields. In Fig. 7 
(b), we have studied the case when the system is experiencing a transverse 
field in addition to the fourth order term. In the presence of the leaking 
term $-g H_x \hat{S}_x$, the population distribution changes completely. Here 
we see transitions between states with $\Delta m = \pm 1$. The population of 
the $M_s = -10$ state decreases drastically from the first plateau to the 
third plateau. We do not see any weight for $M_s = 10 $ state in this case 
also.

In Figs. 8 (a) - (c), we show the time evolution of the magnetization at a 
constant axial field equal to $0.74 ~J/\hbar$. The oscillations are most 
complex when we have both a transverse field and a fourth order term as in 
Fig. 8 (b). The oscillations are somewhat simpler when we have only the fourth 
order anisotropy term and no transverse field as in Fig. 8 (a); we see a beat 
structure superimposed on regular oscillations. Finally, when we only have a 
small transverse field equal to $0.05$ and no fourth order term, the 
magnetization evolves as a smooth cosine function with time as shown in Fig. 
8 (c); the behavior is just like that of a two-level system. Note also that 
the ranges of the magnetization oscillations change considerably with the 
sweep rate in all the three figures.
 
Finally, in Fig. 9, we present the weights of the different $M_s$ states when
there is a constant axial field equal to $0.74$ and a transverse field equal
to $0.05$, but no fourth order term. The weights are dramatically different 
for the two sweep rates; for the lower sweep rate shown in the lower panel,
the system has sufficient time to reach states with higher magnetization.

\section{\bf Summary and Outlook}

We have studied the effects of the sweep rate, fourth order anisotropic term 
and a transverse magnetic field on the quantum tunneling of magnetization in 
$Mn_{12}Ac$. An effort to reproduce the experimental details will require the 
introduction of terms which take into account the other degrees 
of freedom necessary for the spin system to relax back to the ground state 
after a scattering event. It would also be worthwhile to 
extend this approach to finite temperatures to investigate the thermodynamic 
properties of the molecule and to study the phenomenon of thermally assisted 
tunneling \cite{garanin}. For models of the Ising kind, a 
nontrivial response of the magnetization is known
for an alternating field \cite{raedt2}. A study of the same in our model is 
currently in progress. In conclusion, we would like to say that incorporating 
the spin-phonon \cite{bellessa}, hyperfine \cite{friedman2} and dipolar 
interactions \cite{thomas} to quantitatively explain the experiments will be 
a challenge for the future.

\vskip 1 true cm
\leftline{\bf Acknowledgments}
\vskip .5 true cm

We thank Prof. Indrani Bose for introducing us to this problem, and the 
Council of Scientific and Industrial Research, India for their grant No.
01(1595)/99/EMR-II.

\newpage

\begin{center}
\begin{tabular}{|c|c|}

\hline $\hspace{0.2cm}M_{s}$ value \hspace{0.2cm}& \hspace{.5cm}
Dimension of Hilbert Space\hspace{0.5cm} \\

\hline \hspace{0.2cm}$\pm 13$\hspace{0.2cm} & \hspace{0.5cm}
139672\hspace{0.5cm} \\     
\hline \hspace{0.2cm}$\pm 12$\hspace{0.2cm} & \hspace{0.5cm}
269148\hspace{0.5cm} \\     
\hline \hspace{0.2cm}$\pm 11$\hspace{0.2cm} & \hspace{0.5cm}
484144\hspace{0.5cm}  \\    
\hline\hspace{0.2cm}$\pm 10$\hspace{0.2cm} & \hspace{0.5cm}
817176\hspace{0.5cm}  \\   
\hline \hspace{0.2cm}$\pm 9$\hspace{0.2cm} & \hspace{0.5cm}
1299632\hspace{0.5cm} \\    
\hline\hspace{0.2cm} $\pm 8 $\hspace{0.2cm} &\hspace{0.5cm}
1954108\hspace{0.5cm} \\
\hline \hspace{0.2cm} $\pm 7 $\hspace{0.2cm} &\hspace{0.5cm}
2785384\hspace{0.5cm}  \\   
\hline \hspace{0.2cm} $\pm 6\hspace{0.2cm} $ &\hspace{0.5cm}
3772176\hspace{0.5cm}  \\   
\hline\hspace{0.2cm}$\pm 5 $\hspace{0.2cm} &\hspace{0.5cm}
4862352\hspace{0.5cm}  \\   
\hline\hspace{0.2cm} $\pm 4 $\hspace{0.2cm} &\hspace{0.5cm}
5974048\hspace{0.5cm}  \\   
\hline\hspace{0.2cm} $\pm 3 $\hspace{0.2cm} &\hspace{0.5cm}
7003944\hspace{0.5cm}  \\   
\hline\hspace{0.2cm}$\pm 2 $\hspace{0.2cm} &\hspace{0.5cm}
7842070\hspace{0.5cm}  \\
\hline\hspace{0.2cm}$\pm 1 $\hspace{0.2cm} &\hspace{0.5cm}
8390440\hspace{0.5cm}  \\   
\hline\hspace{0.2cm} $ 0 $\hspace{0.2cm} & \hspace{0.5cm}
8581300\hspace{0.5cm} \\ \hline   
\end{tabular}
\vskip 0.5 true cm

\noindent Table 1. Different $M_s$ values and the corresponding Hilbert space 
dimensions. 

\vskip 3.0 true cm
\begin{tabular}{|c|c|}
\hline $S_{total}$ & Energies \\ \hline
  \hspace{0.2cm} 10 \hspace{0.2cm} & \hspace{0.5cm} -53.251824
\hspace{0.5cm} \\ \hline
  \hspace{0.2cm} 11\hspace{0.2cm}  &\hspace{0.5cm}  -52.419914
\hspace{0.5cm} \\ \hline
   \hspace{0.2cm} 12\hspace{0.2cm}  & \hspace{0.5cm} -51.512498
\hspace{0.5cm} \\ \hline
  \hspace{0.2cm} 10\hspace{0.2cm}  &\hspace{0.5cm}  -50.910154
\hspace{0.5cm} \\ \hline
  \hspace{0.2cm} 13\hspace{0.2cm}  & \hspace{0.5cm} -50.529748
\hspace{0.5cm} \\ \hline
\end{tabular}   
\vskip 0.5 true cm
\noindent Table 2. Energies (in units of J) of the lowest five states and the 
corresponding $S_{total}$ values for the Hamiltonian in Eq. (\ref{ham1}).

\end{center}

\newpage

\noindent {\bf Figure Captions}
\vskip 1 true cm

\noindent {1.} A schematic diagram of the exchange interactions between
the $Mn$ ions in the $Mn_{12}Ac$ molecule.

\noindent {2.} Energy spectrum of the Hamiltonian in Eq. (\ref{ham2}) in the 
presence of a time-independent axial field. Only a few low-lying energy 
levels are shown.

\noindent {3.} Evolution of magnetization at two different sweep rates. 
The upper curve is for a time step equal to $0.1$ (in units of 
$\hbar /J$), while the lower curve is for a time step of $0.01$.
The inset shows the result obtained when the field is held fixed for a time
equal to $300 ~\hbar /J$.

\noindent {4.} Magnetization vs. the axial field for two different sweep rates
when there is both a fourth order term and a transverse field equal to 
$0.01 ~J/\hbar$.

\noindent {5.} Magnetization vs. the axial field for two different sweep rates
when there is only a transverse field equal to $0.05$ and no fourth order 
term.

\noindent {6.} Energy level distribution of the system for two different axial 
fields. The two figures on the left have a fourth order term and no 
transverse field, while the two figures on the right have a transverse field
and no fourth order term.

\noindent {7.} Weights of different $M_s$ states at the onset fields of 
three different plateaus is shown in plot (a), and in the presence of an 
additional small transverse field $H_x =0.01$ is shown in plot (b). The fourth 
order term is present in both cases. Note that $\Delta M_s$ is equal to 
$4$ and $1$ in cases (a) and (b) respectively. 

\noindent {8.} Evolution of magnetization with time is shown at a constant 
value of the axial field equal to $0.74$. In (a) there is only a fourth 
order anisotropy term; in (b) there is both a fourth order term and a 
transverse field equal to $0.01$; in (c) there is only a transverse field 
equal to $0.05$. Note that the ranges of the oscillations change considerably 
with the sweep rate.

\noindent {9.} Weights of different $M_s$ states in the presence of an axial 
field equal to $0.74$ and a transverse field equal to $0.05$ for two different 
sweep rates. 

\newpage

\begin{figure}
\begin{center}
\epsfig{file=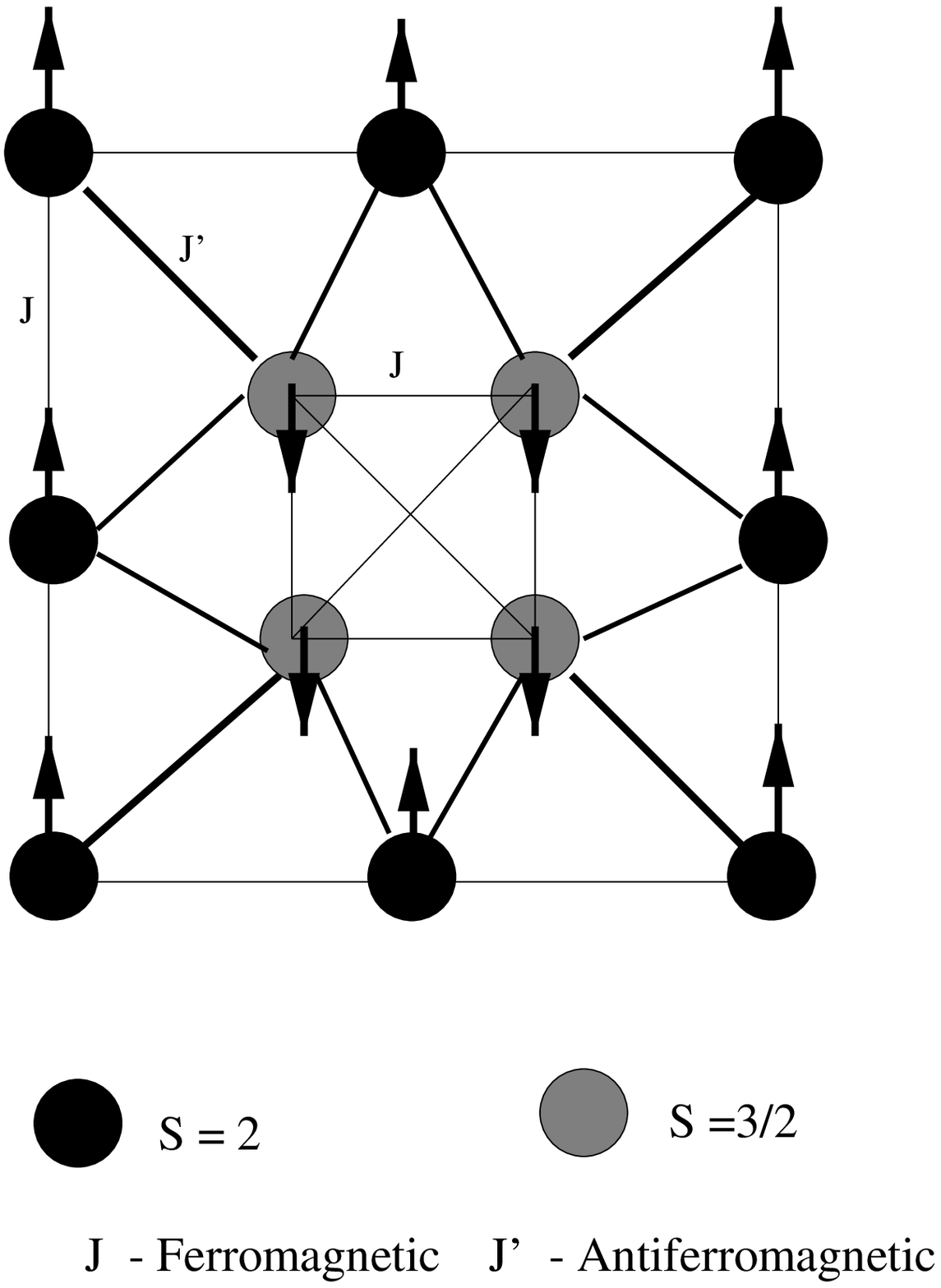}
\end{center}
\vspace*{.4 cm}
\centerline{Fig. 1}
\end{figure}

\begin{figure}
\begin{center}
\epsfig{file=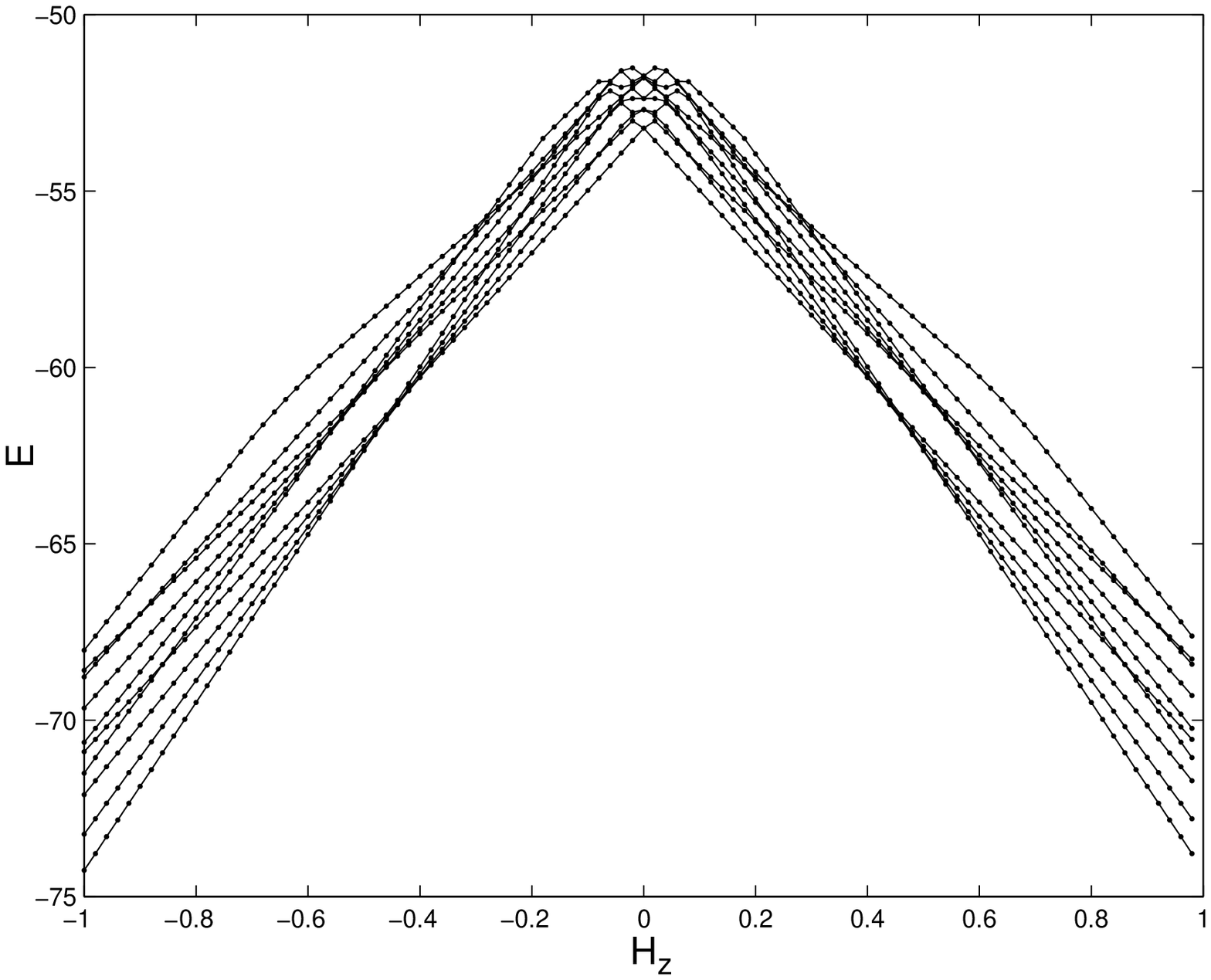}
\end{center}
\vspace*{2 cm}
\centerline{Fig. 2}
\end{figure}

\begin{figure}
\begin{center}
\epsfig{figure=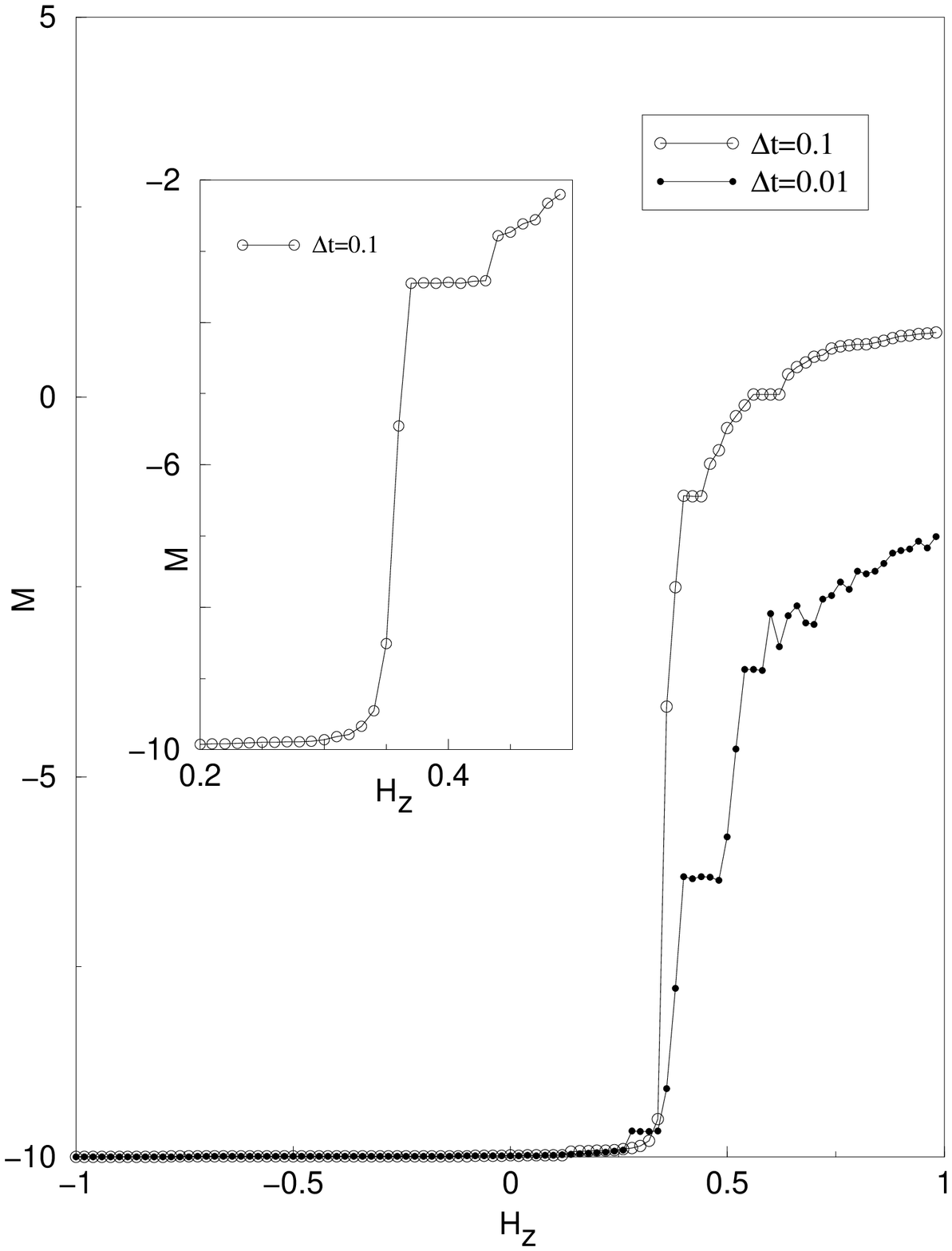,bbllx=50,bblly=0,bburx=500,bbury=800,height=20cm}
\end{center}
\vspace*{-2 cm}
\centerline{Fig. 3}
\end{figure}

\begin{figure}
\begin{center}
\epsfig{figure=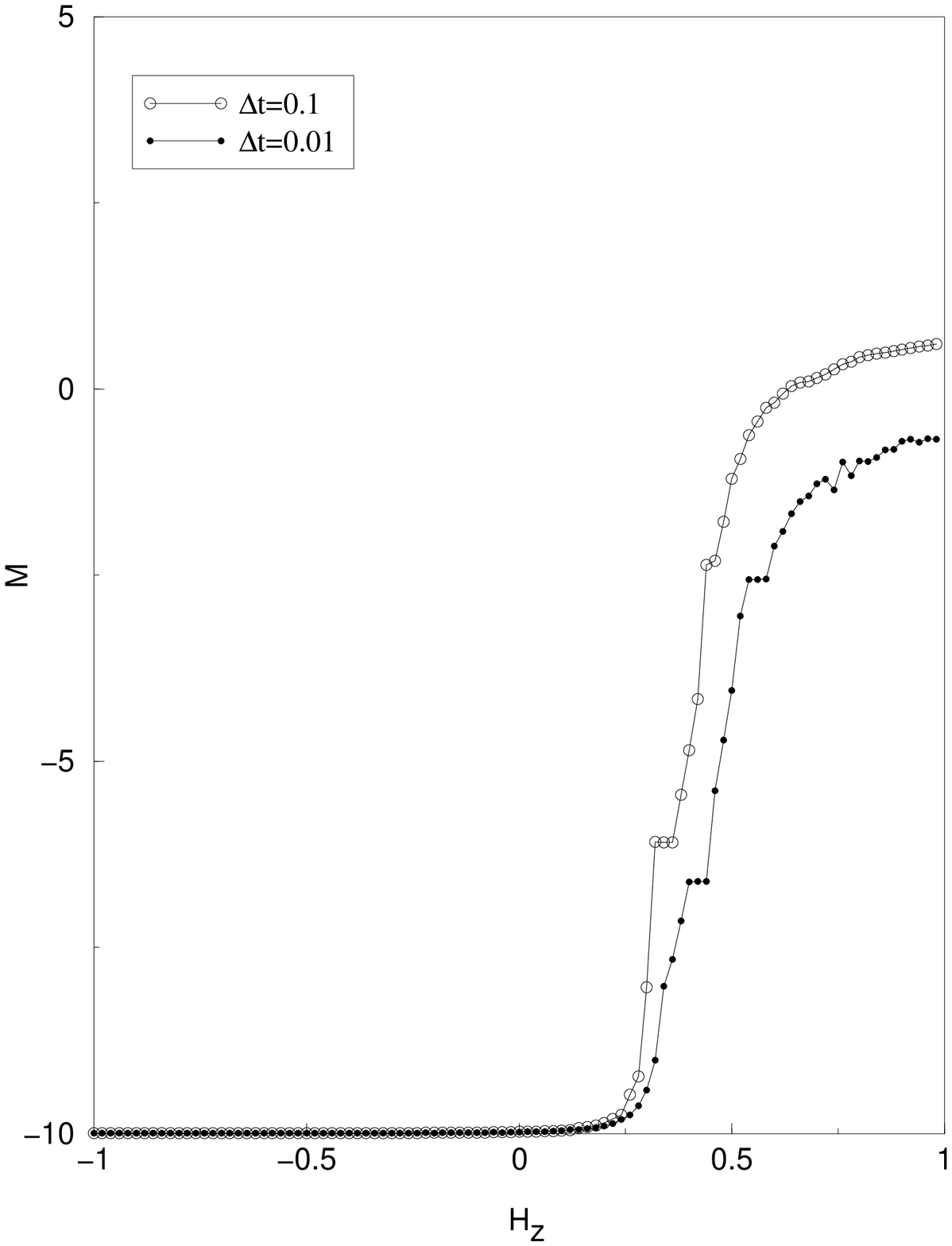,bbllx=50,bblly=0,bburx=500,bbury=800,height=20cm}
\end{center}
\vspace*{-2 cm}
\centerline{Fig. 4}
\end{figure}

\begin{figure}
\begin{center}
\epsfig{figure=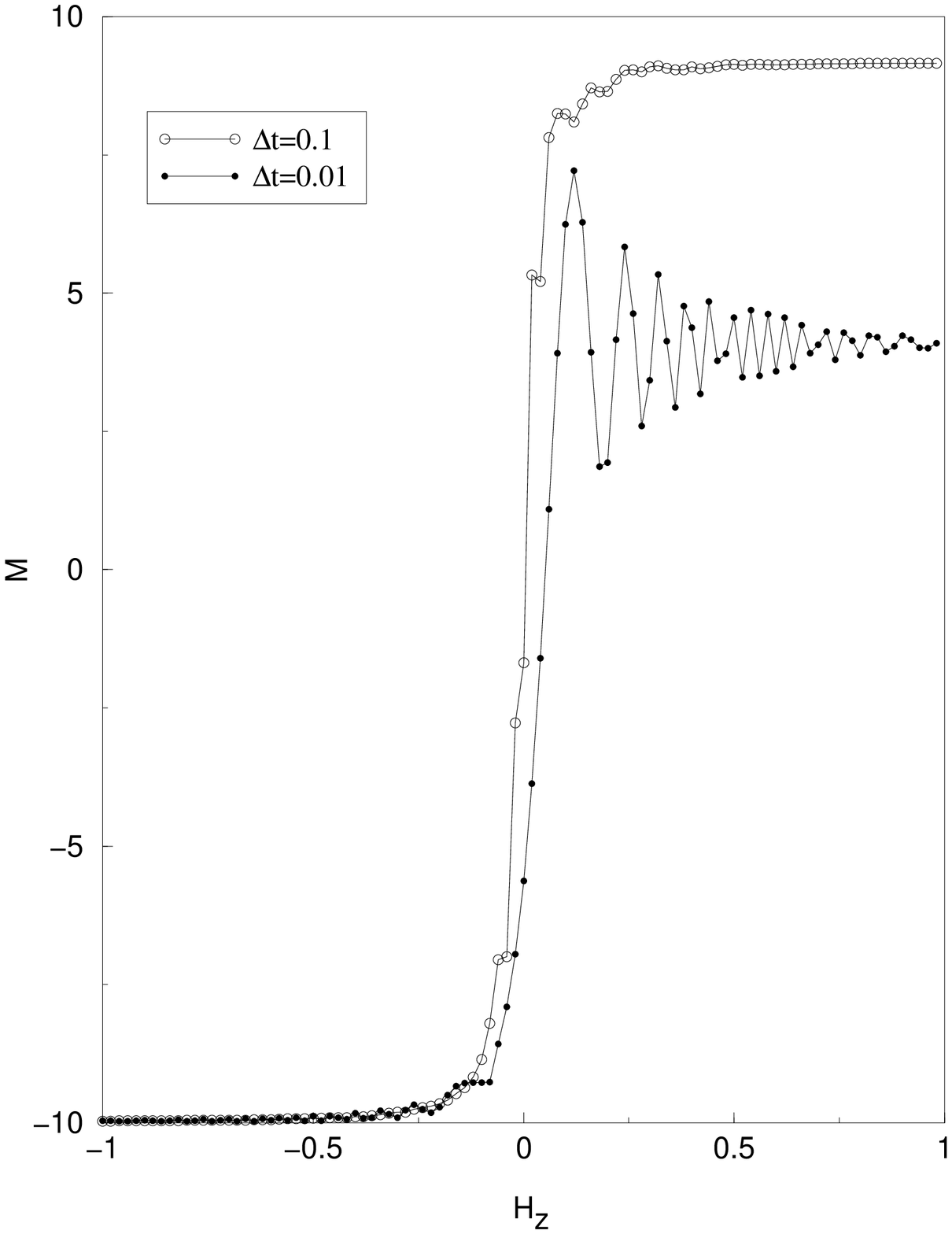,bbllx=50,bblly=0,bburx=500,bbury=800,height=20cm}
\end{center}
\vspace*{-2 cm}
\centerline{Fig. 5}
\end{figure}

\begin{figure}
\begin{center}
\epsfig{figure=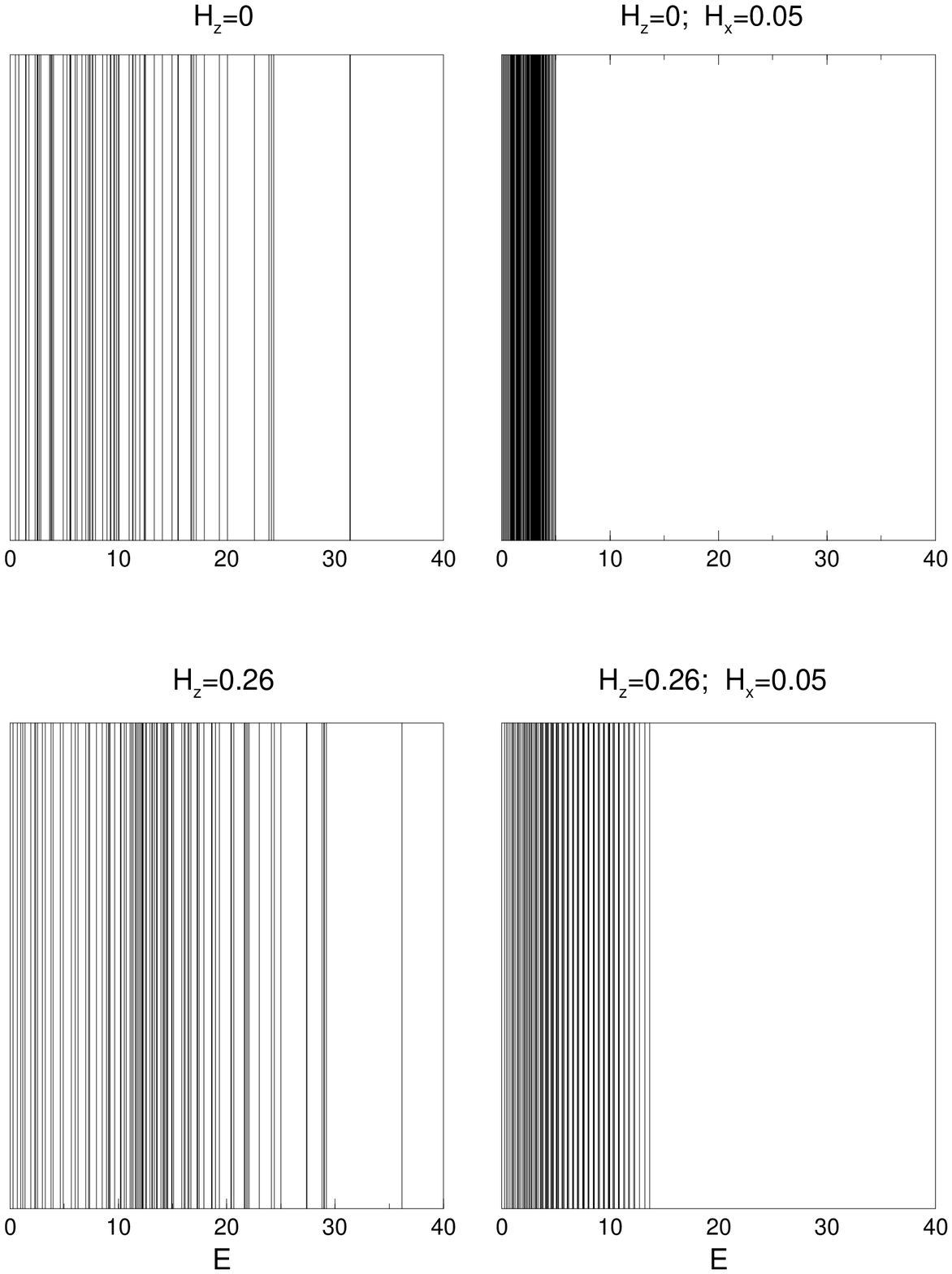,bbllx=50,bblly=0,bburx=500,bbury=800,height=20cm}
\end{center}
\vspace*{-1.6 cm}
\centerline{Fig. 6}
\end{figure}

\begin{figure}
\begin{center}
\epsfig{file=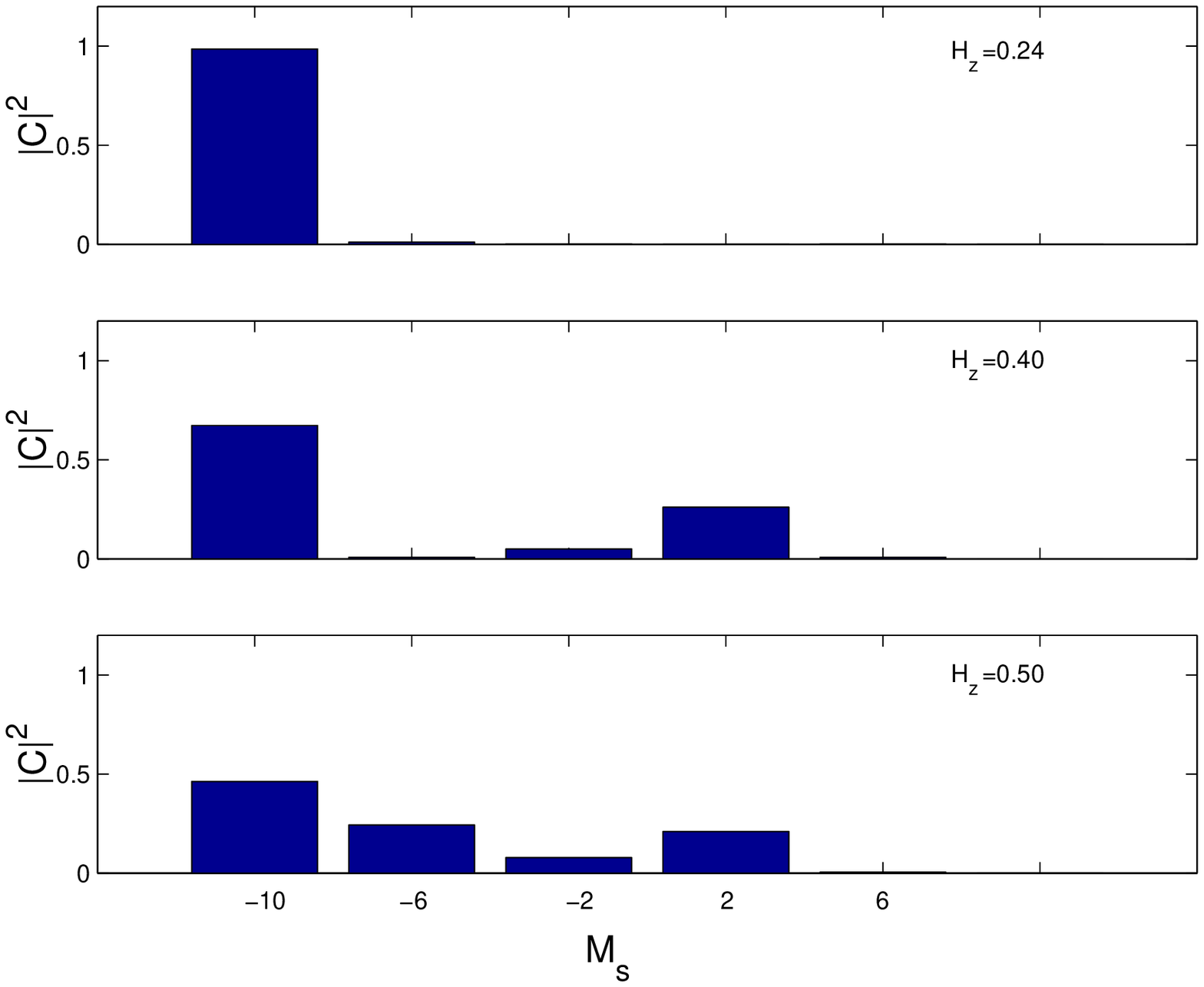}
\end{center}
\vspace*{2 cm}
\centerline{Fig. 7 (a)}
\end{figure}

\begin{figure}
\begin{center}
\epsfig{file=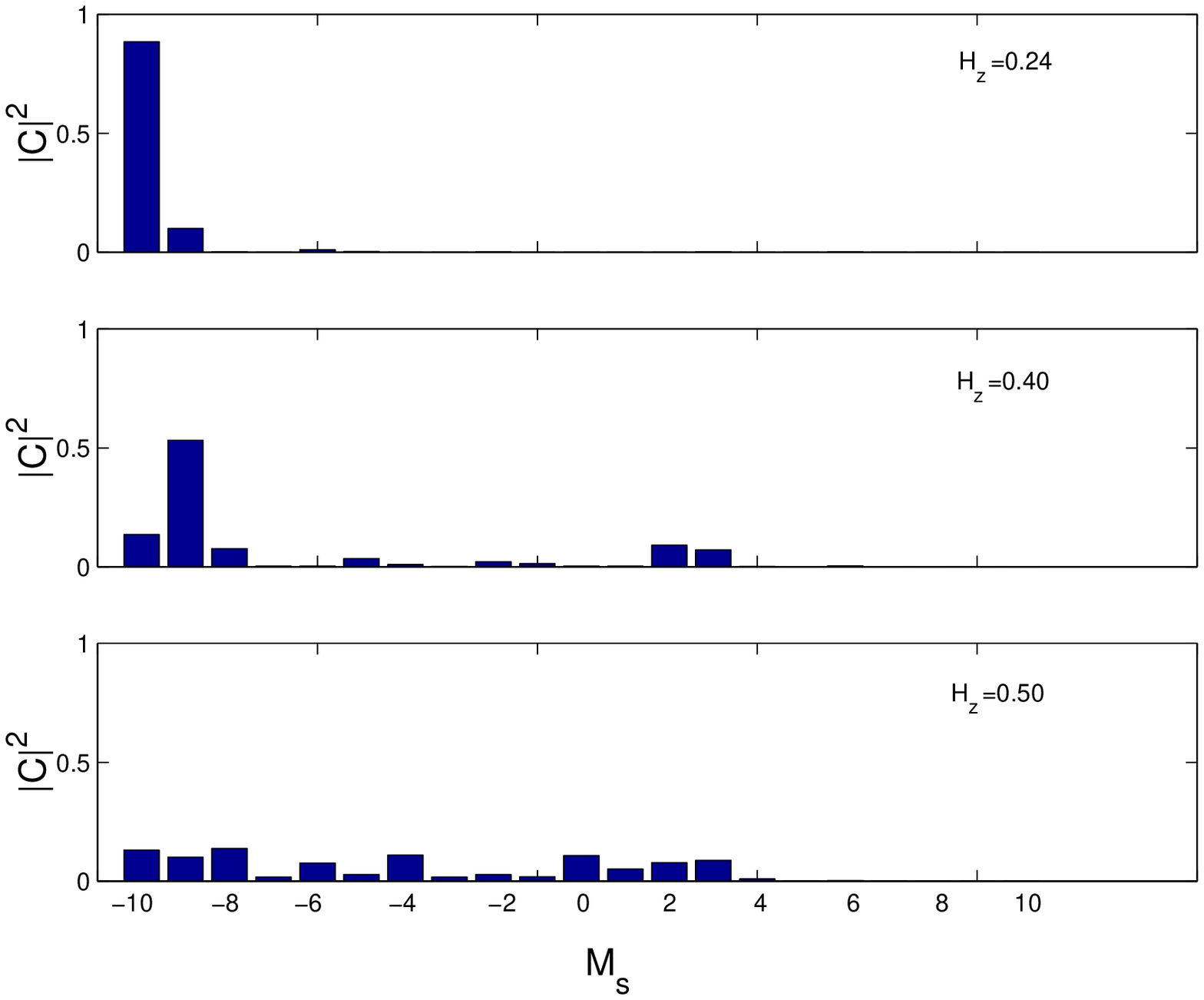}
\end{center}
\vspace*{2 cm}
\centerline{Fig. 7 (b)}
\end{figure}

\begin{figure}
\begin{center}
\epsfig{figure=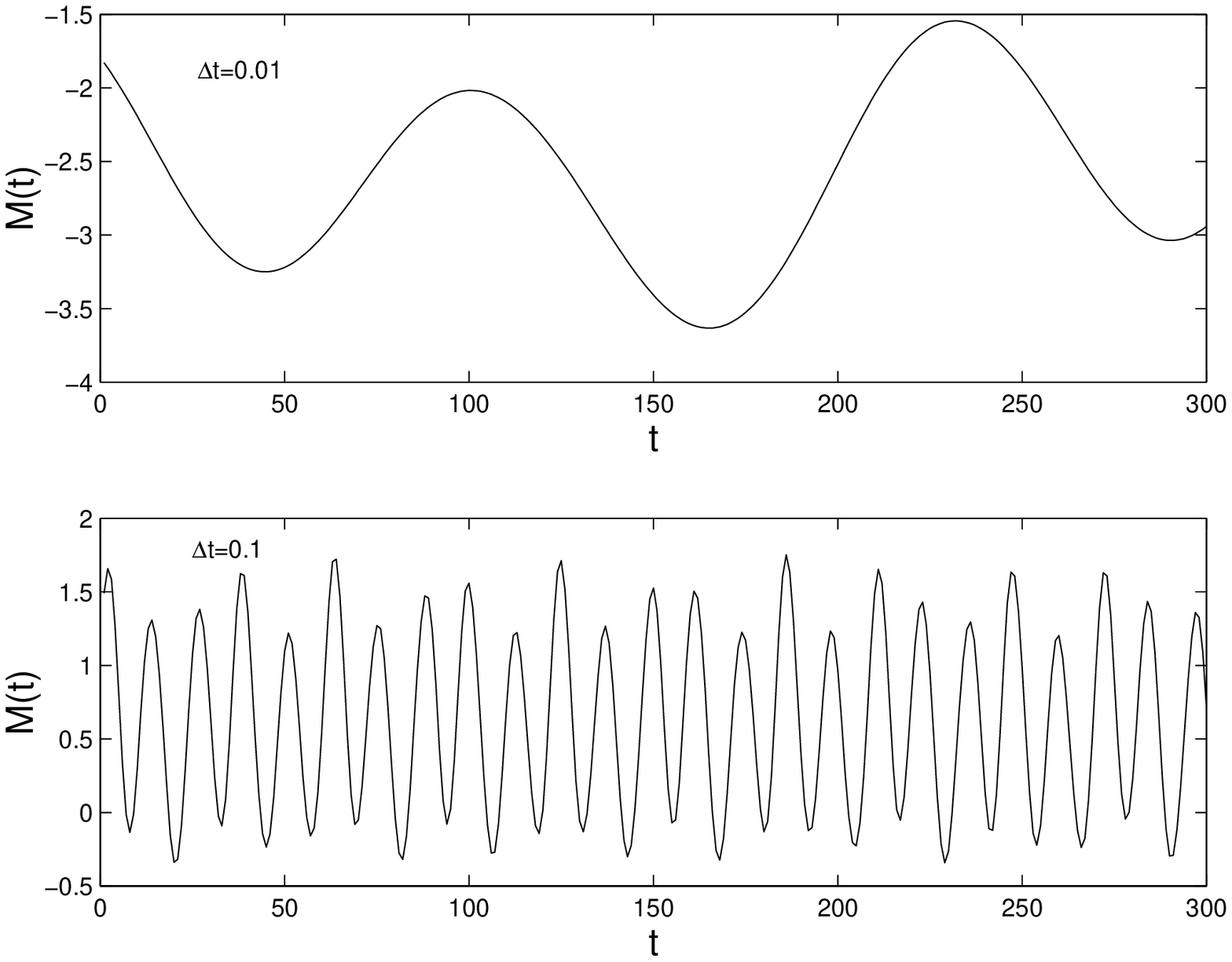,bbllx=50,bblly=0,bburx=500,bbury=800,height=22cm}
\end{center}
\vspace*{-4.4 cm}
\centerline{Fig. 8 (a)}
\end{figure}

\begin{figure}
\begin{center}
\epsfig{figure=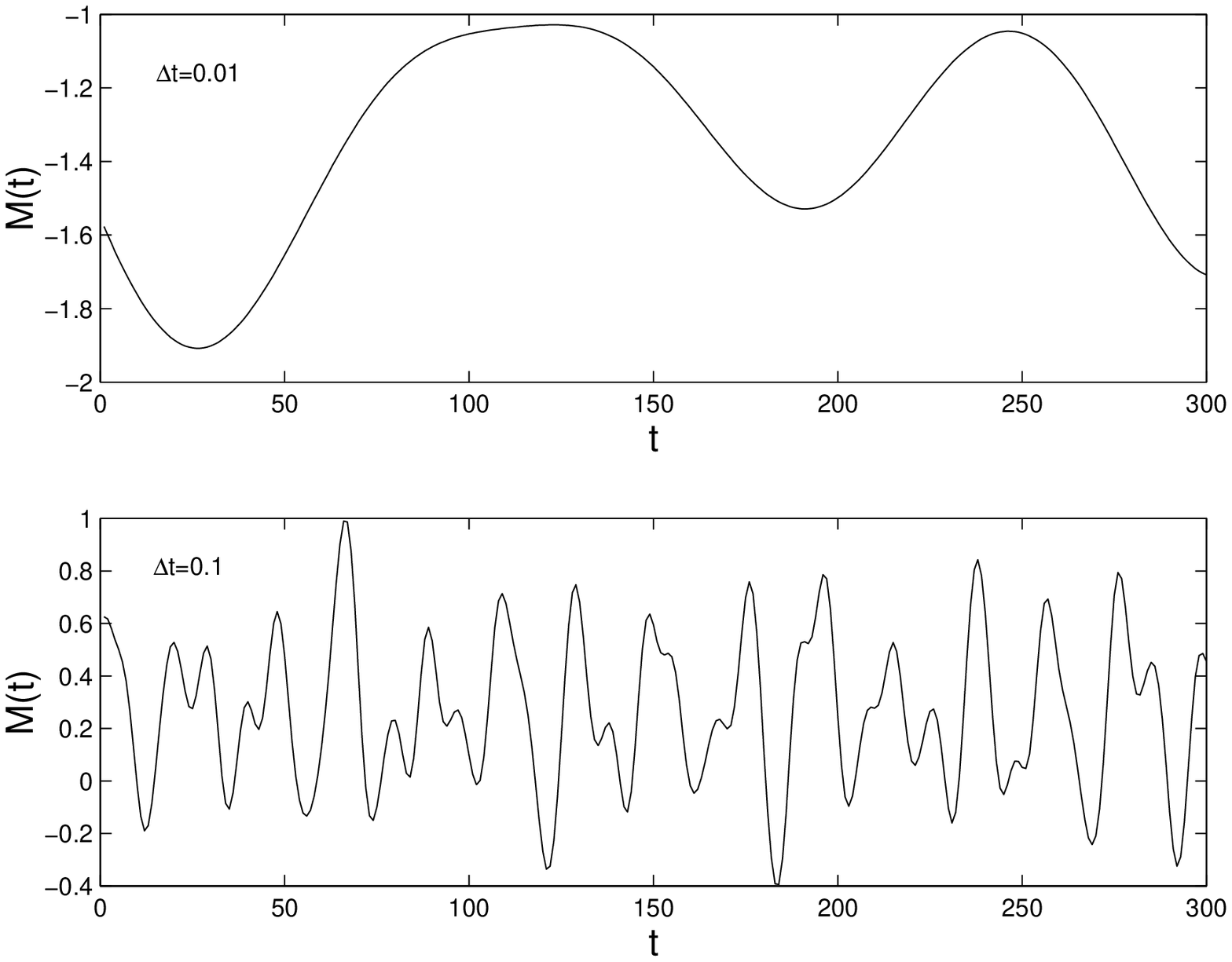,bbllx=50,bblly=0,bburx=500,bbury=800,height=22cm}
\end{center}
\vspace*{-4.4 cm}
\centerline{Fig. 8 (b)}
\end{figure}

\begin{figure}
\begin{center}
\epsfig{figure=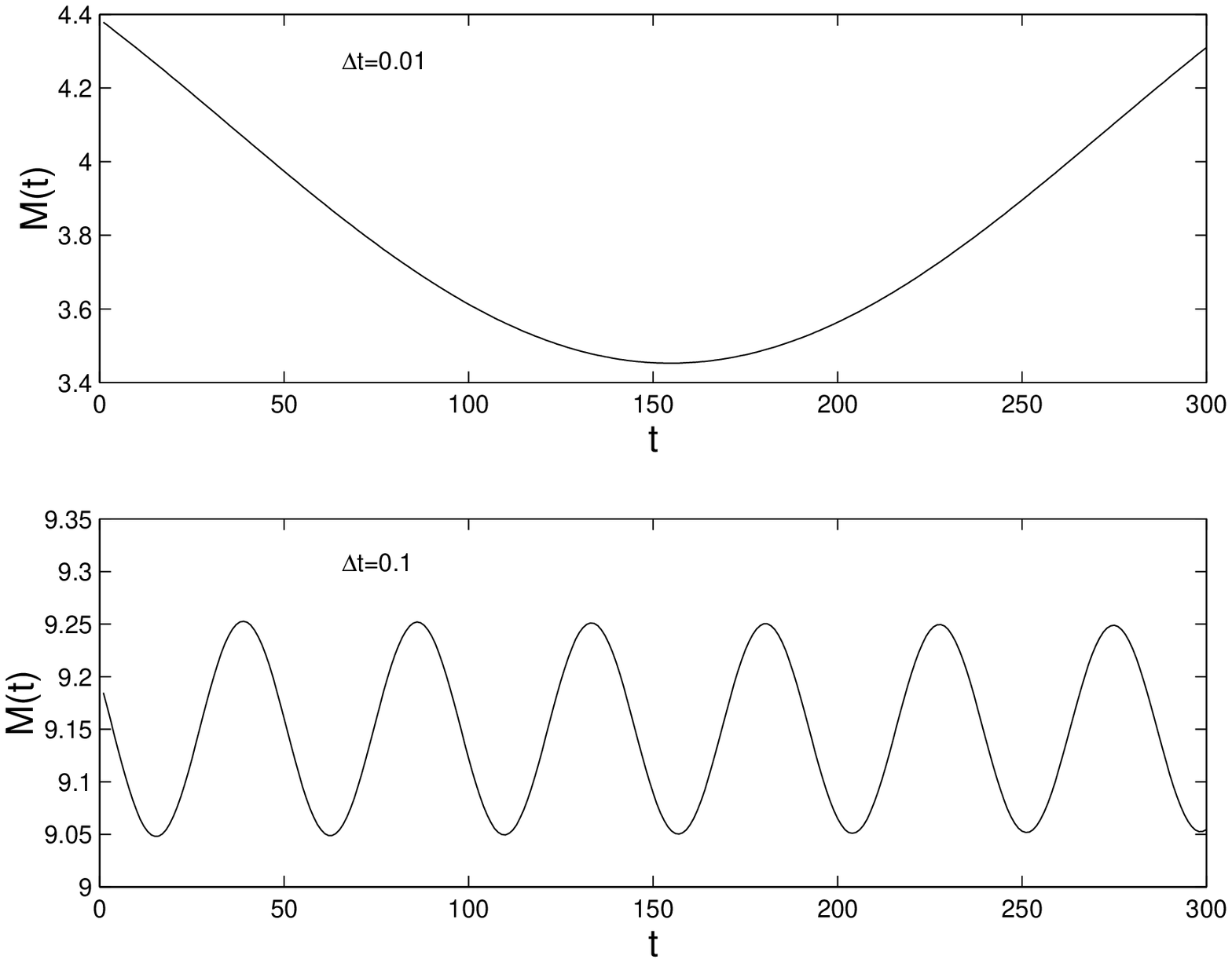,bbllx=50,bblly=0,bburx=500,bbury=800,height=22cm}
\end{center}
\vspace*{-4.4 cm}
\centerline{Fig. 8 (c)}
\end{figure}

\begin{figure}
\begin{center}
\epsfig{file=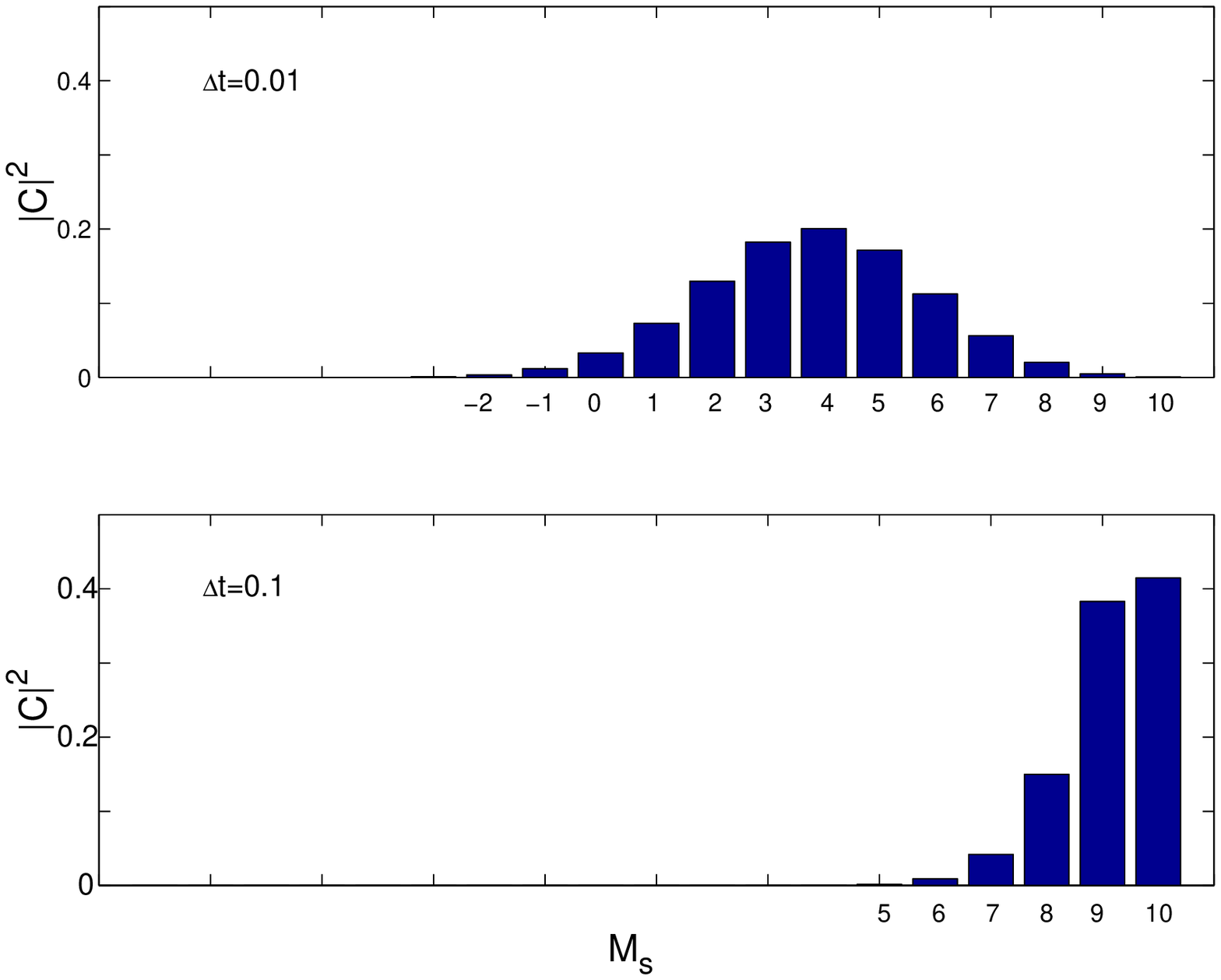}
\end{center}
\vspace*{2 cm}
\centerline{Fig. 9}
\end{figure}

\end{document}